\documentclass[aps,floatfix,prb,twocolumn,showpacs]{revtex4}
\usepackage{amsmath}
\usepackage[final,dvips]{graphicx}
\usepackage{amsfonts}
\usepackage{amssymb}
\begin{document}
\title{Theory of fluctuations in a two-band superconductor}
\author{A.\ E.\ Koshelev$^{1}$, A.\ A.\ Varlamov$^{1,2}$, and V.\ M.\ Vinokur$^{1}$}
\affiliation{$^1$Materials Science Division, Argonne National
Laboratory, 9700 S.Cass Avenue, Argonne Il 60439,\\
$^2$Coherentia-INFM, via del Politecnico, 1, 00133 Rome, Italy}
\date{\today}
\begin{abstract}
A theory of fluctuations in two-band superconductor MgB$_{2}$ is
developed. Since the standard Ginzburg-Landau (GL) approach fails in
description of its properties, we generalize it basing on the microscopic
theory of a two-band superconductor. Calculating the microscopic
fluctuation propagator, we build up the nonlocal two-band GL functional
and the corresponding time-dependent GL equations. This allows us to
calculate the main fluctuation observables such as fluctuation specific
heat and conductivity.
\end{abstract}
\maketitle

\section{Introduction}

The celebrated phenomenological Ginzburg-Landau (GL) theory of
superconductivity was introduced in the seminal paper by Ginzburg and Landau
\cite{GL50} and has been proving ever since to be one of most fruitful and
universal tools in the description basic properties of superconductors.  The
examples of its success range from the prediction of the Abrikosov vortex
state \cite{AbrikosovJETP57} to the recent advance in understanding the
complex vortex phase diagram of the high-temperature superconductors
\cite{BGFLV94} and the properties of mesoscopic superconducting systems
\cite{F03}. The Gor'kov's derivation of the GL equations from the BCS theory
\cite{G58} that have put the GL theory on a firm microscopic basis and
related phenomenological constants to material parameters, completed the
construction of the GL theory of superconductors.

The GL theory describes  well the properties of almost all superconductors
near transition temperature and is successful even when dealing with the
superconductors with quite complicated band structures. The notorious
exception from the rule is recently discovered multiband superconductor
magnesium diboride (MgB$_{2}$).  As it was shown \cite{GK03a,GK03b}, the GL
theory applies to MgB$_2$ only in the very immediate vicinity of the
transition temperature, $T_{c}$, i.e. within the interval which turns out to
be much more narrow than the usual one given by the condition $|T-T_{c}|\ll
T_{c}$. The origin of this narrowing of the interval of validity lies in the
sophisticated band structure of the material and reflects the specific
interplay between its single-electron and superconducting characteristics.
Namely, the Brillouin zone of MgB$_2$ consists of the two families of bands:
the quasi-two-dimensional $\sigma$-bands with the strong superconductivity
and the weakly superconducting three-dimensional $\pi$-bands.  Due to the
large difference in anisotropy, the c-axis coherence length of the
$\pi$-bands, $\xi_{2,z}$, is much larger than the c-axis coherence length of
the $\sigma$-bands, $\xi_{1,z}$.  Formation of the global coherent
superconductivity with the unique order parameter implies the appearance of
the associated unique effective coherence length $\widetilde{\xi}_{z}\left(
T\right)$, which turns out to be much smaller than $\xi_{2,z}$ at almost any
temperature.  As a consequence, the GL applicability interval shrinks to the
parametrically narrow range of temperatures where $\widetilde{\xi}_{z}\left(
T\right)  \gg\xi_{2,z}$. Beyond this temperature range the system is
strongly nonlocal along the $z$-axis; to describe such a nonlocality one has
to employ a generalized nonlocal GL model \cite{GK03b}. One of the
spectacular manifestations of the non-GL behavior in MgB$_2$ is the strong
temperature dependence of the $H_{c2}$ anisotropy close to $T_{c}$\
\cite{hc2anisotropy}.

The description of superconducting fluctuations is one of the
major fields for the application of the GL theory. Since the
standard GL approach turns out to be insufficient for MgB$_2$, a
generalization of the GL theory is required in order to describe
its fluctuation properties. The variation of the order parameter
on the scales smaller than the largest intrinsic coherence length
means that the usually assumed local approximation does not hold
anymore and the corresponding short-wavelength fluctuations have
to be taken into account.

In the present paper we develop a nonlocal theory of
superconducting fluctuations that applies to a strongly
anisotropic two-band superconductor \cite{Multi-Two} and show that
the short-wavelength fluctuations are essential in it near the
critical temperature. We start with the derivation of the
microscopic fluctuation propagator for a such two-band model. Then
we use it to nonlocal two-band GL functional and the corresponding
time-dependent GL (TDGL) equations. This, in particular, allows us
to calculate the kinetic and thermodynamic observable quantities
including the fluctuational specific heat and conductivity.

As we have already stated, the main source of the non-GL behavior is the
nonlocality in the $c$ direction, i.e., the strong wave-vector dependence of
c-axis coherence length. The conventional local form of the GL equations
turns out to be valid only within the narrow interval of temperatures,
$|T-T_c|/T_c\ll \xi_{1,z}^2/\xi_{2,z}^2+S_{12}\ll 1$, where $S_{12}\ll 1$ is
the relative interband interaction constant which will be specified below.
Beyond this interval, the superconducting correlations in the $\pi$-band
become nonlocal and their contribution to the effective coherence length
rapidly decreases. Far away from $T_c$ the effective c-axis coherence length
is determined only by the $\sigma$-band. In other words, the effective
c-axis coherence diverges for $T\rightarrow T_c$ faster than it could be
expected from the naive GL extrapolation, $\widetilde{\xi}_z(T)\propto
1/\sqrt{T-T_c}$, started from high temperatures. This also leads to the
decrease of the effective anisotropy factor $\Gamma
(T)=\widetilde{\xi}_x(T)/\widetilde{\xi}_z(T)$. As a consequence, the
temperatures dependencies of all fluctuation corrections exhibit the
characteristic crossovers between the dominating $\sigma$-band regime (far
away from $T_c$) and the ``true'' GL regime (very close to $T_c$). For
example, the $c$-axis component of the paraconductivity diverges faster in
the immediate vicinity of $T_c$ than one could expect from high-temperature
extrapolation using the Aslamazov-Larkin formula \cite{LV02} while the
fluctuation specific heat and  $ab$ component of the paraconductivity
diverge slower than the corresponding extrapolations. We will obtain the
temperature dependencies of these fluctuation corrections.

\section{Critical temperature and fluctuation propagator}

\subsection{Cooper pairing in two-band model}

The BCS theory was generalized to the case of the two-band electron spectrum
long time ago \cite{Suhl,Moskal} and has been extended recently to include
the specific features of magnesium diboride in Refs.\ \onlinecite
{Kortus,An,LiuPRL01,Kong,Yildirim,Choi}. We briefly overview this theory
rewriting it in terms of Green's function formalism. The two-band BCS
Hamiltonian is given by
\begin{widetext}
\begin{equation}
\mathcal{H}=\sum_{\mathbf{p},\alpha,\sigma}\zeta_{\alpha}\left(
\mathbf{p}\right) \psi_{\alpha,\mathbf{p},\sigma}^{\dag}\psi_{\alpha
,\mathbf{p},\sigma}-\sum_{\mathbf{p,p}^{\prime},\mathbf{q},\alpha,\beta
,\sigma,\sigma^{\prime}}g_{\alpha\beta}\psi_{\alpha,\mathbf{p+q,}\sigma}%
^{\dag}\psi_{\alpha,-\mathbf{p},-\sigma}^{\dag}\psi_{\beta,-\mathbf{p}^{\prime
},-\sigma^{\prime}}\psi_{\beta,\mathbf{p}^{\prime}+\mathbf{q,}%
\sigma^{\prime}}, \label{BCSham}%
\end{equation}
\end{widetext}
where $\psi_{\alpha,\mathbf{p},\sigma}^{\dag}$ and
$\psi_{\alpha,\mathbf{p},\sigma}$ are the creation and annihilation field
operators in the Heisenberg representation for quasiparticle in band
$\alpha$ with momentum $\mathbf{p}$ and spin $\sigma$,
\begin{equation}
\zeta_{\alpha}\left(  \mathbf{p}\right)  =\mathbf{v}_{\alpha}\left(
\mathbf{p}-\mathbf{p}_{F\alpha}\right)  \label{spectrum}%
\end{equation}
is the quasiparticle spectrum, $\mathbf{v}_{\alpha}$,
$\mathbf{p}_{F\alpha}$ are the Fermi velocity and momentum of the
$\alpha$-band. The matrix nature of the electron-electron interaction
$-g_{\alpha\beta}$ in (\ref{BCSham}) reflects the possibility of the
interband interactions. The free electron Green's functions for each band
have the usual form $G_{\alpha}\left(
r,r^{\prime},\tau,\tau^{\prime}\right) =-i\left\langle
T_{\tau}\psi_{\alpha}\psi_{\alpha}^{\prime\dag}\right\rangle $, where
$T_{\tau}$ is the time-ordering operator. In the Matsubara representation,
\begin{equation}
G_{\alpha}\left(  \mathbf{p},\varepsilon_{n}\right)  =\left(  i\varepsilon
_{n}-\zeta_{\alpha}\left(  \mathbf{p}\right)  \right)  ^{-1} \label{green}%
\end{equation}
with $\varepsilon_{n}=2\pi T(n+1/2)$ being the fermionic Matsubara frequencies.

\begin{figure}[ptb]
\begin{center}
\includegraphics[width=3.4in]{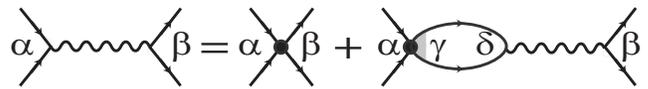}
\end{center}
\caption{Graphic representation of Dyson equation (\ref{Dyson1}) for
the fluctuation propagator $L_{\alpha\beta}$ (wavy line). The Greek
letters indicate the band indices. The black dot represents the
coupling-constants matrix $-g_{\alpha\beta}$, the loop represents
the polarization operator $\Pi_{\gamma\delta}$, and the shaded
triangle represents the scattering vertex
$C_{\gamma\gamma}$.}%
\label{Fig-Dyson}%
\end{figure}

Now we turn to calculation of the fluctuation propagator
$L_{\alpha\beta}$ which characterizes the properties of fluctuation
Cooper pairs and their effect on observable quantities of
superconductor above $T_{c}$.\cite{LV02} Note first of all, that the
interband electron interactions do not result in the Cooper pairing
of the electrons from different bands but rather lead to the
transfer of the pairing correlations between the bands. Indeed, the
Cooper pairing means the appearance of superconducting-type
correlations between two similar states obeying the condition of the
time reversal symmetry. This means that pairing is possible only for
electrons belonging to the same band, otherwise the electron states
are too diverse (in terms of the plane waves description, their
momenta are not the opposite) and the integral of the product of
their wave functions is zero. Thus the formation of unique
condensate should be understood as the result of the
\textit{intraband} electron correlations and subsequent
\textquotedblleft travel\textquotedblright of the Cooper pair from
one band to another due to off-diagonal interaction components
$g_{12}$ and $g_{21}$. Hence, in terms of diagrams, the entrance and
exit lines of the fluctuation propagator must belong to the same
bands (see Fig.\ \ref{Fig-Dyson}) and it can be presented as the
$2\times2$ matrix $L_{\alpha\beta}\left(
\mathbf{q},\Omega_{k}\right) $ where $\Omega_{k}\!=2\pi Tk$ are the
bosonic Matsubara frequencies. The corresponding Dyson equation for
fluctuation propagator can be written in the ladder approximation as
(see Fig.\ \ref{Fig-Dyson})
\begin{equation}
L_{\alpha\beta}=-g_{\alpha\beta}+g_{\alpha\gamma}\Pi_{\gamma\delta}%
L_{\delta\beta}, \label{Dyson1}%
\end{equation}
where $\Pi_{\alpha\beta}\left(  \mathbf{q},\Omega_{k}\right)  $ is the
matrix polarization operator which is determined by two Green's functions
loop \cite{LV02}. Without interband electron scattering the only nonzero
components of the operator $L_{\alpha\beta}$ are the diagonal ones.

Turning to the role of electron scattering, let us note that at the first
sight the short-range impurity potential might lead to the scattering of
electrons all over the whole Fermi surface which would result in the
equalization of all gap values and reduction of $T_{c}$. However, it is
well established now that this does not happen in MgB$_{2}$: even in the
samples with rather strong intraband scattering, interband scattering
remains rather weak. This property gave possibility to fabricate samples
with very high upper critical fields (up to 50 tesla) which have
transition temperatures only slightly smaller than clean
material\cite{highHc2}. Weakness of the interband scattering  has been
explained by Mazin \emph{et al.} in Ref.\ \onlinecite{M03} who argued that
different parity symmetry of the $\sigma$- and $\pi$- orbitals leads to
strong reduction of the impurity scattering matrix element between these
bands. For simplicity, we will completely neglect the interband
scattering. In this case the off-diagonal components of the polarization
operator vanish. The averaging over impurities position will result just
in the usual renormalization of the Green's functions:
$\varepsilon_{n}\rightarrow \varepsilon_{n}+1/2\tau_{\alpha}$
($\tau_{\alpha}$\ is the corresponding intraband scattering time) and to
the appearance of the scattering vertex part $C_{\alpha\alpha}$ in the
expression for the polarization operator
\begin{widetext}
\begin{equation}
\Pi_{\alpha\alpha}\left(  \mathbf{q},\Omega_{k}\right)  =T\sum_{\varepsilon
_{n}}\int\frac{d\mathbf{p}}{\left(  2\pi\right)  ^{3}}C_{\alpha\alpha
}(\mathbf{q},\varepsilon_{n+k},-\varepsilon_{n})G_{\alpha}\left(
\mathbf{p+q},\varepsilon_{n}+\Omega_{k}\right)  G_{\alpha}\left(
-\mathbf{p},-\varepsilon_{n}\right)  . \label{polarizator}%
\end{equation}
\end{widetext}
The scattering vertex part can be calculated in the ladder
approximation\cite{AGD}. In the case of a dirty metal $\left(  T\tau
_{\alpha\alpha}\ll1\right)  $ this gives\cite{LV02}
\begin{equation}
C_{\alpha\alpha}(\mathbf{q},\varepsilon_{n+k},-\varepsilon_{n}) \!=\!\frac
{1/\tau_{\alpha\alpha}}{|\varepsilon_{n}\!+\!\varepsilon_{n+k}| \!+\!\hat
{D}_{\alpha}q^{2}}\theta\left(  \varepsilon_{n+k}\varepsilon_{n}\right)  ,
\label{cooperon}%
\end{equation}
leading to the following result for the polarization operator
\begin{equation}
\Pi_{\alpha\alpha}(\mathbf{q},\Omega_{k}) \!=\!\nu_{\alpha}\!\left[  \ln\!
\frac{\omega_{D}}{2\pi T}\!-\!\psi\!\left(  \frac{1}{2}\!+\!\frac{|\Omega
_{k}|\!+\!\hat{D}_{\alpha}q^{2}}{4\pi T}\right)  \right]  , \label{polar_expl}%
\end{equation}
where $\nu_{\alpha}$ is the density of states in the band $\alpha$,
$\omega_{D}$ is the Debye frequency and $\psi\left(  x\right)  $ is
the digamma-function. We introduced above the notation for the
diffusion-coefficient tensor acting in the momentum and band spaces:
\begin{equation}
\hat{D}_{\alpha}q^{2}\equiv \sum_{a}D_{\alpha,a}q_{a}^2.
\label{diffusion}%
\end{equation}
The diffusivities $D_{\alpha,a}$ determine the band coherence lengths as
$\xi_{\alpha,a}^{2}=\pi D_{\alpha ,a}/8T$. For magnesium diboride the
ratio
\begin{equation}
r=D_{2,z} /D_{1,z}=\xi_{2,z} ^{2}/\xi_{1,z}^{2}, \label{ratio}
\end{equation}
is large, $r\gg 1$, due to the large difference between the band
Fermi velocities in the $c$ direction \cite{Kortus,An}. This
parameter will play an important role in the following
consideration.

The inversion of the Dyson equation (\ref{Dyson1}) gives
\begin{equation}
\widehat{L}^{-1}=-\widehat{\emph{g}}^{-1}+\widehat{\Pi}. \label{propa}%
\end{equation}
We will use now this general expression to reconsider the main properties of
two-band superconductivity and to describe the corresponding fluctuation properties.

\subsection{Critical temperature}

The critical temperature of a two-band superconductor is determined by the
condition $\det\widehat{L}^{-1}=0$ taken at $\Omega_{k}=0$ and $\mathbf{q}%
=0$:
\[
\left(  \frac{g_{22}}{\det\widehat{\emph{g}}}\!-\!\nu_{1}\ln\frac{2\gamma_{E}
\omega_{D}}{\pi T_{c}}\right)  \!\! \left(  \frac{g_{11}}{\det\widehat
{\emph{g}} }\!-\!\nu_{2}\ln\frac{2\gamma_{E}\omega_{D}}{\pi T_{c}}\right)
\!-\!\frac{g_{21}g_{12} }{\det^{2}\widehat{\emph{g}}}\!=\!0.
\]
with $\ln\gamma _{E}\equiv C_{E} \approx0.577$ being the Euler constant.
Introducing the coupling-constants matrix as
\[
\widehat{\lambda}=\left(
\begin{array}
[c]{cc}%
\nu_{1}g_{11} & \nu_{2}g_{12}\\
\nu_{1}g_{21} & \nu_{2}g_{22}%
\end{array}
\right)  ,
\]
we can find that the transition temperature is determined by its largest
eigenvalue,
\[
\tilde{\lambda}=\frac{\lambda_{+}}{2}+\sqrt{\frac{\lambda_{-}^{2}}{4}
+\lambda_{12}\lambda_{21}}
\]
with $\lambda_{\pm}\equiv\lambda_{11}\pm\lambda_{22}$, and it is given by the
BCS-type equation:
\[
\ln\frac{2\gamma_{E}\omega_{D}}{\pi T_{c}}=\tilde{\lambda}^{-1}.
\]

Following Ref.\ \onlinecite{GK03b}, we introduce the inverse
coupling-constants matrix%
\begin{widetext}
\[
\widehat{W}    =\left(
\begin{array}
[c]{cc}%
W_{11} & -W_{12}\\
-W_{21} & W_{22}%
\end{array}
\right)  \equiv\widehat{\lambda}^{-1}-\tilde{\lambda}^{-1}\hat{I}
=\frac{1}{\det\widehat{\lambda}}\left(
\begin{array}
[c]{cc}%
\!\sqrt{  \lambda_{-}^{2}/4\!+\!\lambda
_{12}\lambda_{21}}\!-\!\lambda_{-}/2 & -\lambda_{21}\\
-\lambda_{12} & \!\sqrt{\lambda_{-}^{2}/4
\!+\!\lambda_{12}\lambda_{21}}\!+\!\lambda_{-}/2%
\end{array}
\right) ,
\]
\end{widetext}
where $\hat{I}$ is the unit matrix and $\det\widehat{\lambda}\equiv
\lambda_{11}\lambda_{22}-\lambda_{12}\lambda_{21}$. From the definition of
the matrix $\widehat{W}$ is evident that it is degenerate,
$\det\widehat{W}\equiv W_{11}W_{22}-W_{12}W_{21}=0.$ Having in mind
applications of our theory to MgB$_{2}$, we will use the numerically
computed effective coupling constants for this compound \cite{Golub}:
\begin{equation}
\left(
\begin{array}
[c]{cc}%
W_{11} & W_{12}\\
W_{21} & W_{22}%
\end{array}
\right)  =\left(
\begin{array}
[c]{cc}%
0.088 & 0.535\\
0.424 & 2.56
\end{array}
\right)  .
\end{equation}
We see now that apart from the already mentioned large ratio $r$
(\ref{ratio}), another characteristic small parameter, the relative
interband coupling,
\begin{equation}
S_{12}=\frac{W_{11}}{W_{22}}\approx0.035\ll1
\end{equation}
appears\cite{GK03a,GK03b}. In what follows we will demonstrate that
it is the interplay between these two parameters that defines the
rich picture of \ fluctuations that occurs beyond the traditional GL
theory.

\subsection{Fluctuation propagator}

In terms of the notations  introduced above, the inverse matrix for
the
propagator (\ref{propa}) assumes the form%
\begin{widetext}
\begin{equation}
\widehat{L}^{-1}(\Omega_{k},\mathbf{q})=-\left(
\begin{array}[c]{cc}
W_{11}+\epsilon+\beta_{1}\left( |\Omega_{k}|,q\right)&-W_{12}\\
-W_{21} & W_{22}+\epsilon+\beta_{2}\left( |\Omega_{k}|,q\right)
\end{array}
\right)  \left(
\begin{array}[c]{cc}
\nu_{1} & 0\\
0 & \nu_{2}%
\end{array}
\right)  , \label{linvers}%
\end{equation}
where $\epsilon=\ln (T/T_{c})$,
\[
\beta_{\alpha}\left( \omega,q\right)\equiv
\beta\left[  \frac{\pi}{8 T} (\omega +\hat{D}_{\alpha}q^{2})\right] ,
\]
and
\begin{equation}
\beta(x) \equiv\psi\left(  1/2+2x/\pi^{2}\right)  -\psi\left(
1/2\right)=\left\{
\begin{tabular}
[c]{l}%
$x,\ \ x\ll1$\\
$\ln x+C_{E}+\ln(8/\pi^{2}),\ x\gg1$%
\end{tabular}
\right., \label{beta1}
\end{equation}
Note that due to the identity $W_{12}\nu_{2}=W_{21}\nu_{1}$, the matrix
$\widehat{L}^{-1}(\Omega_{k},\mathbf{q})$ is symmetric.

Calculating $\det\widehat{L}^{-1}$ and inverting the matrix, one can
present $\widehat{L}(\Omega _{k},\mathbf{q})$ as
\begin{equation}
\widehat{L}(\Omega_{k},\mathbf{q})=\frac{\left(
\begin{array}
[c]{cc}%
\widetilde{W}_{22}+\widetilde{\theta}\left[  \epsilon+
\beta_{2}\left(|\Omega_{k}|,q\right) \right] &
\widetilde{W}_{21}\\
\widetilde{W}_{12} & \widetilde{W}_{11}+\widetilde{\theta}\left[
\epsilon+\beta_{1}\left( |\Omega_{k}|,q\right)\right]
\end{array}
\right)  \left(
\begin{array}
[c]{cc}%
\nu_{1}^{-1} & 0\\
0 & \nu_{2}^{-1}%
\end{array}
\right)  }{\epsilon+\left[  \widetilde{W}_{11}+\widetilde{\theta}%
\epsilon\right] \beta_{2}\left(|\Omega_{k}|,q\right)+\left[  \widetilde{W}_{22}+\widetilde{\theta}%
\beta_{2}\left(|\Omega_{k}|,q\right)\right]
\beta_{1}\left(|\Omega_{k}|,q\right)}, \label{propagator1}
\end{equation}
where we introduced the following notations
\[
\widetilde{W}_{ik}=\frac{W_{ik}}{W_{11}+W_{22}},\widetilde{\theta}=\frac
{1}{W_{11}+W_{22}}.
\]
Since we are interested in the most singular contribution with respect to
$\epsilon$, we omitted the terms $\epsilon^{2}$ and $\epsilon$ as compared
to $W_{22}$.

In order to derive microscopically the TDGL equation and study the effect of
fluctuations on transport phenomena one has to perform the analytical
continuation of the propagator (\ref{propagator1}) from the set of the
imaginary Matsubara frequencies $i\Omega_k=2\pi i T k$, $k=0,1,2\ldots$ to
the upper half-plane of the complex frequencies. Since the function
$\psi(z)$ has poles at points $z_n = 0, -1, -2\ldots$ the analytical
continuation to the real frequencies $\Omega$ can be performed by the simple
substitution $\Omega_k \rightarrow-i\Omega$ in the argument of the function
(\ref{beta1}).\cite{E61,LV02} When this argument is small ($\Omega,
D_{\alpha}q^2 \ll T$) the function $\beta$ is reduced to the Fourier
transform of the diffusion operator:
\[
\beta\left[  \frac{\pi}{8 T} (|\Omega_{k}|+\hat{D}_{\alpha}q^{2})\right]
\rightarrow-i\frac{\pi\Omega}{8T}+\frac{\pi\hat{D}_{\alpha}q^{2}}{8T}.
\]
Performing the analytical continuation, using
$\widetilde{W}_{22}\approx1$, and accounting for the small parameters of
the model ($\widetilde{W}_{11}\approx
S_{12},\widetilde{\theta},\epsilon,\Omega/T, \hat{D}_{1}q^2/T\ll1$), we
obtain the analytically continued propagator
$\widehat{L}^{R}(\mathbf{q,}\Omega)$ in the form
\begin{equation}
\widehat{L}^{R}(\mathbf{q,}\Omega)\approx-\frac{1}{\epsilon+\beta_{1}\left(-i\Omega,q\right)
+\frac{S_{12}\beta_{2}\left(-i\Omega,q\right)}{1+\widetilde{\theta}\beta_{2}\left(-i\Omega,q\right)}}\left(
\begin{array}
[c]{cc}%
1/\nu_{1} & \frac{\widetilde{W}_{21}}{\nu_{2}\left(  1+\widetilde{\theta}%
\beta_{2}\left(-i\Omega,q\right)\right)  }\\
\frac{\widetilde{W}_{12}}{\nu_{1}\left(
1+\widetilde{\theta}\beta_{2}\left(-i\Omega,q\right)\right)
} & \frac{S_{12}+\widetilde{\theta}\left[  \epsilon+\beta_{1}\left(-i\Omega,q\right)\right]  }%
{\nu_{2}\left(  1+\widetilde{\theta}\beta_{2}\left(-i\Omega,q\right)\right)  }%
\end{array}
\right)  . \label{progen1}%
\end{equation}
\end{widetext}
Below we will analyze the behavior of this matrix in different temperature
intervals.

\subsubsection{Ginzburg-Landau regime}

Let us take the reduced temperature so small that both $\beta_{1}$ and
$\beta_{2}$ functions can be expanded. At the end of this subsection we will
arrive at the analytical criterion for this condition to be satisfied. In
this case the propagator (\ref{progen1}) significantly simplifies (we use
here the definition $\hat{D}_{\alpha}q^{2}/4\pi T=2/\pi^{2}\xi
_{\alpha,a}^{2}q_{a}^{2}$):
\begin{equation}
\widehat{L}^{R}\left( \mathbf{q},\Omega\right)
\!=\!-\frac{1}{\epsilon\!-\!i\gamma_{GL}
\Omega\!+\!\widetilde{\xi}_{a}^{2}q_{a}^{2}}\left(
\begin{array}
[c]{cc}%
1/\nu_{1} & \widetilde{W}_{21}/\nu_{2}\\
\widetilde{W}_{12}/\nu_{1} & S_{12}/\nu_{2}%
\end{array}
\right)  , \label{proGL}%
\end{equation}
where $\gamma_{GL}=\pi/8T$, and the effective coherence length
components $\widetilde{\xi}_{a}$ is given by
\begin{equation}
\widetilde{\xi}_{a}^{2}=\widetilde{W}_{11}\xi_{2,a}^{2}+\widetilde{W}_{22}%
\xi_{1,a}^{2}\approx\xi_{1,a}^{2}\left(  1+S_{12}\frac{\xi_{2,a}^{2}}%
{\xi_{1,a}^{2}}\right)  ,\label{xiGL}%
\end{equation}
Typically $\xi_{2,x}\approx\xi_{1,x}$ and, due to the presence of a
small coefficient $S_{12}$, the contribution from the $\pi$-band to
the transversal component of the effective coherence length \
($a=x,y$) turns out to be small and can be ignored. On the other
hand, the c-axis motion in the $\pi$-band, due to the large value of
the ratio $r=\xi_{2,z}^{2}/\xi_{1,z}^{2}$ can play an important role
for the longitudinal component of the effective coherence length
\[
\widetilde{\xi}_{z}^{2}=\xi_{1,z}^{2}\left(  1+S_{12}r\right)
\]
and when $S_{12}\gg 1/r$ it may significantly exceed $\xi_{1,z}^{2}$.

The characteristic momentum $q_{z}$ is determined by the diverging GL
coherence length:
\begin{equation}
q_{z}\sim\sqrt{\epsilon}/\widetilde{\xi}_{z}.\label{char}%
\end{equation}
Therefore the standard GL theory is valid only at temperatures so
close to $T_{c}$ that $\xi_{2,z}^{2}q_{z}^{2}\ll1$ and where both
$\beta$ functions can be expanded. The corresponding condition
acquires the form \cite{GK03a,GK03b}
\begin{equation}
\epsilon\ll\frac{\xi_{1,z}^{2}}{\xi_{2,z}^{2}}+S_{12}=1/r+S_{12}.
\label{GLcondition}
\end{equation}

\subsubsection{Beyond the GL regime}

Let us return to the propagator (\ref{propagator1}) and analyze its
behavior in all the vicinity of the transition, $\epsilon\ll1$, performing
allowed expansions and simplifications. The diffusion in the first band is
slow, therefore we can expand the corresponding $\beta$-functions. On the
other hand, the function $\beta_{2}(-i\Omega,\mathbf{q})\equiv\beta\left[
\frac{\pi}{8 T}(-i\Omega+\hat{D}_{2}q^{2})\right]  $ can be expanded only
with respect to $\Omega$ and $q_{\Vert}$ and one has to keep the full
nonlinear dependence on $q_{z}$. As a result, we obtain the expression
similar to (\ref{proGL}) except for the appearance of the explicit $q_{z}%
$-dependence of the GL coefficients $\gamma_{GL}\rightarrow\gamma\left(
q_{z}\right)  $ and $\widetilde{\xi}_{a}^{2}\left(  q_{z}\right)  $
\begin{widetext}
\begin{equation}
\widehat{L}^{R}(\mathbf{q,}\Omega)=-\frac{1}{\epsilon-i\gamma\left(
q_{z}\right)  \Omega+\widetilde{\xi}_{a}^{2}\left(  q_{z}\right)  q_{a}^{2}%
}\left(
\begin{array}
[c]{cc}%
1/\nu_{1} & \frac{\widetilde{W}_{21}}{\nu_{2}\left(  1+\widetilde{\theta}%
\beta_{2}\right)  }\\
\frac{\widetilde{W}_{12}}{\nu_{1}\left(
1+\widetilde{\theta}\beta_{2}\right)  } &
\frac{S_{12}+\epsilon\widetilde{\theta}+\widetilde{\theta}\xi_{1,a}^{2}%
q_{a}^{2}}{\nu_{2}\left(  1+\widetilde{\theta}\beta_{2}\right)  }%
\end{array}
\right)  . \label{Lgeneral}%
\end{equation}
\end{widetext}
with
\begin{subequations}
\begin{align}
\gamma\left(  q_{z}\right)   &  \approx\gamma_{GL}\left[  1+\frac
{S_{12}+\epsilon\widetilde{\theta}}{1+\widetilde{\theta}\beta_{2}(
q_{z})}\beta^{\prime}_2( q_{z})\right],
\label{gammanonGL}\\
\widetilde{\xi}_{x}^{2}\left(  q_{z}\right)   &  =\xi_{1x}^{2}+\frac
{S_{12}+\epsilon\widetilde{\theta}}{1+\widetilde{\theta}\beta_{2}(
q_{z})}\beta^{\prime}_2(
q_{z})%
\xi_{2x}^{2},\label{xinonGL1}\\
\widetilde{\xi}_{z}^{2}\left(  q_{z}\right)   &  =\xi_{1,z}^{2}+\frac{S_{12}%
}{1+\widetilde{\theta}\beta_{2}( q_{z})}\frac{\beta_{2}(
q_{z})}{\xi_{2,z}^{2}q_{z}^{2}}%
\xi_{2,z}^{2}. \label{xinonGL2}%
\end{align}
\end{subequations}
Here
\begin{subequations}
\begin{eqnarray}
&\beta^{\prime}(x)=d\beta(x)/dx, \label{betadef1}\\
&\beta_{2}(q_{z}) \equiv\beta\left(  \xi_{2,z}^{2}q_{z}^{2}\right),\label{betadef2} \\
&\beta^{\prime}_{2}(q_{z}) \equiv \beta^{\prime}\left(
\xi_{2,z}^{2}q_{z}^{2}\right).\label{betadef3}
\end{eqnarray}
\end{subequations}

One can see that when the condition (\ref{GLcondition}) is satisfied, the
characteristic $q_{z}$ (see (\ref{char}) ) is small and the expressions
(\ref{gammanonGL}), (\ref{xinonGL1}), and (\ref{xinonGL2})\ reproduce the
TDGL coefficients. In the region
\begin{equation}
1/r+S_{12}\ll\epsilon\ll1
\end{equation}
the argument $\xi_{2,z}^{2}q_{z}^{2}$ of the $\beta$ function
(\ref{betadef2}) for the essential momenta $q_{z}$ becomes large and it
can not be expanded anymore. This means that the gradient expansion needed
for validity of the GL regime fails.
In this region of temperatures the contribution of
the $\pi$-band rapidly decreases and the main role passes to the
$\sigma$-band. Typically, the $\pi$-band strongly contributes to
$\widetilde{\xi}_{z}^{2}\left(  q_{z}\right)  $ and gives only
small corrections to $\gamma\left(  q_{z}\right)  $ and
$\widetilde{\xi}_{x} ^{2}\left(  q_{z}\right)  $.

Now we have in place all the elements required for the microscopic
calculations of fluctuation effects in a two-band superconductor.
However, because of the complex band structure and necessity to
take into account the short wavelength fluctuations the
diagrammatic calculations present itself a bulky calculus. In what
follows we establish another route to address fluctuation
phenomena. Namely, we will re-derive the GL functional and extend
the standard GL scheme to treatment of fluctuations for the
two-band superconductor, and then apply this modified GL approach
to calculations of specific heat and paraconductivity.

\section{Nonlocal two-band GL functional and TDGL equations}

\subsection{GL functional and TDGL equations}

Knowing the explicit form of the fluctuation propagator (\ref{linvers}), one
can write down corresponding GL free-energy functional $\emph{F}_{GL}%
=\emph{F}_{GL}^{\left(  2\right)  }+\emph{F}_{GL}^{\left(  4\right)  }.$ The
complete procedure of its microscopic derivation is given in Ref.\
\onlinecite{LV02} and here we will present only the specific expressions for
GL\ coefficients corresponding to the model under consideration.

The quadratic in order parameter $\Delta_{\alpha}$ part of the GL functional
$\emph{F}_{GL}^{\left(  2\right)  }$\ is expressed in terms of the linearized
GL Hamiltonian density $H_{\alpha\beta}(\mathbf{q})\equiv L_{\alpha\beta}%
^{-1}\left(  \mathbf{q},\Omega=0\right)  :$
\begin{equation}
\emph{F}_{GL}^{\left(  2\right)  }=\int d\mathbf{r}\Delta_{\alpha}^{\ast
}H_{\alpha\beta}(\mathbf{\hat{q}})\Delta_{\beta},\label{F_GL}%
\end{equation}
with
\begin{equation}
H_{\alpha\beta}=a_{\alpha\beta}+\left(
\begin{array}
[c]{cc}%
\nu_{1}\xi_{1,a}^{2}q_{a}^{2} & 0\\
0 & \nu_{2}\left[    \xi_{2,x}
^{2}q_{\parallel}^{2}\beta^{\prime}_2(q_z)+\beta_2(q_z) \right]
\end{array}
\right)  , \label{HGL}%
\end{equation}
and $\mathbf{\hat{q}}=-i\nabla-\frac{2\pi}{\Phi_{0}}\mathbf{A}%
$. Here the matrix
\begin{equation}
\widehat{a}=\left(
\begin{array}
[c]{cc}%
\nu_{1}\left(  W_{11}+\epsilon\right)   & -\nu_{2}W_{12}\\
-\nu_{1}W_{21} & \nu_{2}\left(  W_{22}+\epsilon\right)
\end{array}
\right)
\end{equation}
plays the role of the GL coefficient $a$. In general, we have to
keep nonlinear dependence on $q_{z}$ in
$H_{\alpha\beta}(\mathbf{q})$ meaning that
$H_{\alpha\beta}(\mathbf{\hat{q}})$ in Eq.\ (\ref{F_GL}) is not a
simple second-order differential operator. Note that the similar
generalization of the GL functional has been performed by Maki
\cite{MakiHc2} in order to describe a dirty superconductor in the
vicinity of the $H_{c2}(T)$ line.

The coefficients in the fourth order term of the GL\ functional $\emph{F}%
_{GL}^{\left(  4\right)  }$ do not change their form with respect to the
noninteracting bands case:%
\begin{equation}
\emph{F}_{GL}^{\left(  4\right)  }=b_{\alpha}\int
d\mathbf{r}|\Delta_{\alpha }|^{4}
\end{equation}
with
\[
b_{\alpha}=7\zeta(3)\nu_{\alpha}/(8\pi^{2}T^{2})=\nu_{\alpha}b_{GL}
\]
and $\zeta(3)\approx1.202$.

Now one can write down the linearized TDGL equation in the form
\cite{LV02}
\[
L_{\alpha\beta}^{-1}\left(  \mathbf{q},\Omega\right)
\Delta_{\beta}\! \equiv\! \left[ -i\Omega\gamma_{\alpha\beta}\left(
q_z\right) \!+\!H_{\alpha\beta}\left( \mathbf{q}\right)  \right]
\Delta_{\beta} \!=\!0
\]
with the matrix of TDGL dynamic coefficients
\begin{equation}
\gamma_{\alpha\beta}(q_z) =\frac{\pi}{8T}\left(
\begin{array}
[c]{cc}%
\nu_{1} & 0\\
0 & \nu_{2}\beta^{\prime}_2(q_z)
\end{array}
\right) \label{gammamat}%
\end{equation}
which directly follows from the dynamic part of the fluctuation
propagator. We see that there are two essential differences from the
standard GL approach: (i) strong nonlocality along $z$-direction
(ii) two-component character of the order parameter. In the case
where the parameter $W_{22}$ is large, one can reduce the functional
with two order parameters to the functional for the single
band-averaged order parameter.\cite{GK03b}

\subsection{Spectral properties of $\hat{L}^{-1}(\mathbf{q},\Omega)$}

The TDGL operator $\widehat{L}^{-1}\left( \mathbf{q},\Omega\right)
=\hat {H}\left( \mathbf{q}\right) -i\Omega\hat{\gamma}\left(  q_z
\right) $ is the $2\times2$ matrix defined on the band index space.
It is convenient to diagonalize it. The eigenvalues
$L_{m}^{-1}(\mathbf{q})=\varepsilon_{m}(\mathbf{q})-i\Omega
\gamma_{m}(q_z)$ and the normalized eigenstates $\psi_{m,\beta}$
obey the equation
\begin{equation}
\left[  H_{\alpha\beta}\left(  \mathbf{q}\right)
-i\Omega\gamma_{\alpha\beta}\left( q_z\right)  \right]
\psi_{m,\beta}=L_{m}^{-1}(\mathbf{q},\Omega)\psi_{m,\alpha},
\label{spectral}%
\end{equation}
where $m=1,2$ is the mode index. Superconducting instability
corresponds to the vanishing of the $m=1$ eigenvalue at
$\epsilon,q,\Omega=0$.

Calculating the determinant of (\ref{spectral}) and equating it to zero one
finds\begin{widetext}
\begin{align*}
L_{1,2}^{-1}\left( \mathbf{q},\Omega\right)   & =\frac{\nu_{1}\left(
W_{11}+h_{1}(\mathbf{q},\Omega)\right)
+\nu_{2}\left(  W_{22}+\beta_{2}(q_z)+h_{2}(\mathbf{q},\Omega)\right)  }{2}\\
&  \pm\sqrt{\left(  \frac{\nu_{1}\left(
W_{11}+h_{1}(\mathbf{q},\Omega)\right) -\nu_{2}\left(
W_{22}+\beta_{2}(q_z)+h_{2}(\mathbf{q},\Omega)\right)  }{2}\right)
^{2}+\nu_{1}\nu_{2}W_{11}W_{22}}
\end{align*}
\end{widetext}
with%
\begin{align*}
h_{1}(\mathbf{q},\Omega)  &
\equiv\epsilon+\frac{\pi i\Omega}{8T}+\xi_{1,a}^{2}q_{a}^{2},\\
h_{2}(\mathbf{q},\Omega)  &
\equiv\epsilon+\beta^{\prime}_2(q_z)\left[\frac{\pi
i\Omega}{8T}+\xi_{2,x}^{2} q_{\parallel}^{2}\right].
\end{align*}

\paragraph{GL regime --}

In the case of small $\Omega$, $\epsilon$ and $q$, corresponding to the GL
region of temperatures, one can find the simplified expressions for the
eigenvalues of energy:
\begin{align}
L_{1}^{-1}(\mathbf{q})  &  \approx\nu_{2}\nu_{1}\frac{\left(  W_{22}+W_{11}\right)  }%
{\nu_{1}W_{11}+\nu_{2}W_{22}}\nonumber\\
&  \times\left[  \epsilon+i\Omega\gamma_{GL}+\widetilde{W}_{22}\xi_{1,a}%
^{2}q_{a}^{2}+\widetilde{W}_{11}\xi_{2,a}^{2}q_{a}^{2}\right]
\label{EigenvalueGL1}\\
L_{2}^{-1}  &  \approx\nu_{1}W_{11}+\nu_{2}W_{22}. \label{EigenvalueGL2}%
\end{align}
The expression (\ref{EigenvalueGL1}) for $L_{1}^{-1}(\mathbf{q})$ reproduces
the GL relation (\ref{proGL}) while $L_{2}^{-1}$ is not singular.

\paragraph{Beyond GL regime --}

In generally, the value of function $\beta_{2}(q_z)$ may not be small. In
the absence of that the smallness the more general expansions for
eigenvalues have to be used:
\begin{align}
L_{1}^{-1}(\mathbf{q})  &  \approx\nu_{1}\nu_{2}\frac{\left(
h_{2}(\mathbf{q})\!+\!\beta_{2}(q_z)\right)\!
W_{11}+\!h_{1}(\mathbf{q})\!\left( W_{22}+\!\beta_{2}(q_z)\right)
}{\nu_{1}W_{11}+\nu_{2}\left(
W_{22}+\beta_{2}(q_z)\right)  }\nonumber\\
&  \approx\nu_{1}\left(  h_{1}(\mathbf{q})+\frac{W_{11}\left(
h_{2}(\mathbf{q})+\beta_{2}(q_z)\right)
}{\left(  W_{22}+\beta_{2}(q_z)\right)  }\right); \label{Eigenvalue1}\\
L_{2}^{-1}(\mathbf{q})  &  \approx\!\frac{\nu_{1}W_{11}(\nu_{2}W_{22}\!+\!\nu_{1}W_{11})}%
{\nu_{2}\left(  W_{22}\!+\!\beta_{2}(q_z)\right)
\!+\!\nu_{1}W_{11}}\!+\!\nu_{2}\left(
W_{22}\!+\!\beta_{2}(q_z)\right) \label{Eigenvalue2}.
\end{align}
We will need further the eigenvector for the singular mode, $m=1$, in the
zero order with respect to the small parameters $h_{\alpha}$%
\begin{equation}
\binom{\psi_{1,1}(q_z)}{\psi_{1,2}(q_z)}\approx\binom{\sqrt{1-a^{2}(q_z)}
}{-a(q_z)}
\label{eigenvector}%
\end{equation}
where
\[
a(q_z)=\frac{\sqrt{\nu_{1}\nu_{2}W_{11}W_{22}}}{ \nu_{2}\left(  W_{22}+\beta
_{2}(q_z)\right)  }.
\]
Now we are prepared to revise the results of the standard
fluctuation theory and to generalize them for the case of a two-band
superconductor.

\section{Fluctuation properties of two-band superconductor}

Before going to detail calculations of different fluctuation properties, it
is instructive to estimate the relative strength of fluctuations in the
available two-band material, MgB$_{2}$. It is characterized by the magnitude
of the Ginzburg-Levanyuk parameter:
\[
Gi=\left(  \frac{4\pi^{2}\lambda_{x}^{2}\Gamma (T_{c}) T_{c}}{\Phi_{0}^{2}\widetilde{\xi_{x}}%
}\right)  ^{2},
\]
where $\Phi_{0}=2.07\cdot10^{-7}$ G cm$^2$ is the flux quantum,
$\lambda_{x}$ and $\widetilde{\xi_x}$ are the in-plane London penetration
depth and coherence length, and $\Gamma(T_c)$ is the anisotropy parameter
in the limit $T\rightarrow T_{c}$. Making use of the values of parameters
typical for the clean MgB$_{2}$ crystals: penetration depth
$\lambda_{x}=10^{-5}$cm (Ref.\ \onlinecite{CarringtonPhysC03}), coherence
length $\xi_{x}=10^{-6}$cm (Ref.\ \onlinecite{hc2anisotropy}) and the
anisotropy coefficient $\Gamma (T_{c})=2.5$ (Ref.\
\onlinecite{hc2anisotropy}), we obtain $Gi\approx 1.5\cdot10^{-6}$, what
means that fluctuations in this compound are weak. This conclusion is not
so surprising, since MgB$_{2}$ is known to be a good metal with large
concentration of charge carriers. Therefore, identifying experimentally
the contribution of fluctuations in the clean MgB$_{2}$ crystals is a
challenging task. On the other hand, the amplitude of fluctuation is
expected to be much higher in disordered films or in crystals with large
number of substitution impurities. We stress, however, that the effects
discussed in this paper hold only as long as scattering does not mix bands
and this limits an applicability of our theory for strongly disordered
materials. Another complication is that increasing disorder in magnesium
diboride is usually accompanied by doping of the $\sigma$-band leading to
modification of material parameters (e.g., decreasing the $\sigma$-band
anisotropy) \cite{Kurtus04}.

\subsection{Specific heat}

We start with the calculation of the fluctuation contribution to the
free energy of the two-band superconductor above the critical
temperature. It is determined by the partition function $Z$:
$F=-T\ln Z$, which, in its turn, can be expressed in terms of the
determinant of the GL matrix Hamiltonian (\ref{HGL})
\begin{align*}
F  &  =-T\ln\int\!\!D\Delta\int\!\!D\Delta^{\ast}\exp\left(
-\frac{1}{T}\int \frac{d^{3}q}{\left(  2\pi\right)
^{3}}H_{\alpha\beta}\Delta_{\alpha}^{\ast
}\Delta_{\beta}\right) \\
&  =-TV\int\frac{d^{3}q}{\left(  2\pi\right) ^{3}}\ln\frac{A}{\det
H_{\alpha\beta}},
\end{align*}
where $A$ is an insignificant dimensional constant. Separating the
most singular fluctuation contribution (see Eq.\ (\ref{Lgeneral})),
one can find
\[
\frac{F_{sng}}{V}\!\approx\!-T\!\int\!\frac{d^{3}q}{(2\pi)^{3}}\ln\frac{A}{\left(
1\!+\!\tilde{\theta}\beta_{2}(q_z)\right)\!  \left(
\epsilon\!+\!\xi_{1,a}^{2}q_{a}
^{2}\!+\!S_{12}\tilde{\beta}_{2}(q_z)\right)  }%
\]
with $\tilde{\beta}_{2}(q_z)\equiv\beta_2(q_z) /\left(
1+\tilde{\theta}\beta_2(q_z) \right) $. The corresponding contribution to
the specific heat is given by
\begin{align}
C^{\prime}  &  =-\frac{T}{V}\frac{d^{2}F_{sng}}{dT^{2}}
\approx\int\frac{d^{3}q}{(2\pi)^{3}}\frac
{1}{\left(  \epsilon+\xi_{1,a}^{2}q_{a}^{2}+S_{12}\tilde{\beta}_{2}(q_z)\right) ^{2}} \nonumber\\
&  =\frac{1}{8\pi^{2}\xi_{1,x}^{2}}\int\limits_{-\infty}^{\infty}
dq_{z}\frac{1}{\epsilon+\xi_{1,z}
^{2}q_{z}^{2}+S_{12}\tilde{\beta}_{2}(q_{z})},
\label{SpecHeat}%
\end{align}
Note that the denominator of the the logarithm argument coincides
with the denominator of the fluctuation propagator (\ref{Lgeneral}).
Introducing the reduced variable $u=\xi_{1,z}q_{z}/\sqrt{\epsilon}$,
we rewrite this result in the form convenient for numerical
evaluation:
\begin{equation}
C^{\prime}=\frac{\kappa(\epsilon)}{8\pi\xi_{1,x}^{2}\xi_{1,z}\sqrt{\epsilon}}
, \label{SpecHeat1}
\end{equation}
where the dimensionless function $\kappa(\epsilon)\equiv
\kappa(\epsilon, r, S_{12},\tilde{\theta})$ is defined as
\begin{equation}
\kappa(\epsilon)= \int\limits_{0}^{\infty}
\frac{2du}{\pi}\frac{1}{1\!+\!u^{2}\!+\!(S_{12}
/\epsilon)\tilde{\beta}_2\left(  r\epsilon u^{2}\right)  }.
\label{kappa}
\end{equation}
It weakly depends on temperature and has the following asymptotics
\begin{widetext}
\[
\kappa(\epsilon)  \approx \left\{
\begin{array}
[c]{ll}%
\left(1+S_{12}r\right)^{-1/2}=\xi_{1,z}/\widetilde{\xi}_{z},&\text{for }\epsilon\ll 1/r+S_{12}\\
\sqrt{1-(S_{12}/\epsilon)\ln\left[ Cr\left( \epsilon-S_{12}
\right)\right]}, &\text{for } 1/r+S_{12}\ll \epsilon\ll S_{12}\ln
(rS_{12})\\
1-\frac{S_{12}}{2\epsilon}\frac{\ln\left(  C_{1}r\epsilon\right)  }%
{1+\tilde{\theta}\ln\left(  C_{1}r\epsilon\right) },&\text{for }S_{12}\ln
(rS_{12})\ll\epsilon\ll1 \label{interpol}
\end{array}
\right.
\]
with $C_{1}=16\pi^{-2}\gamma_{E}\exp\left( -2\right) \approx 0.391$.

The formulas (\ref{SpecHeat1})-(\ref{kappa}) work in the entire
region $\epsilon\ll1$. The intermediate asymptotic in $\kappa$
appears only in the case $rS_{12}\gg1\ $ which is valid for MgB$_2$.
\cite{Kortus} One can see that the main difference between the GL
and non-GL regions in the temperature dependence of the fluctuation
heat capacity correction is the change in the coefficient of the
$\epsilon^{-1/2}$ dependence in factor
$\widetilde{\xi}_{z}/\xi_{1,z}$.

\subsection{Paraconductivity}

The paraconductivity in the phenomenological GL approach can be expressed
via the eigenvalues of the GL Hamiltonian and the matrix
elements of the ``velocity''\ operator\ $\widehat{v}_{\alpha}%
^{a}=\partial H_{\alpha\alpha}\left(  q\right)  /\partial q_{a}$.
The approach of Ref.\ \onlinecite{LV02} can be extended it to the
case of two-band model. The general formula for the
paraconductivity tensor then is:
\begin{align}
\sigma^{ab}  &  =\frac{4Te^{2}}{\hbar}\int\frac{d^{3}\mathbf{q}}{\left(
2\pi\right)  ^{3}}\sum_{n,k}\frac{\widehat{v}_{nk}^{a}\widehat{v}_{kn}%
^{b}\gamma_{k}\gamma_{n}}{\varepsilon_{n}\varepsilon_{k}\left(  \gamma
_{n}\varepsilon_{k}+\varepsilon_{n}\gamma_{k}\right)  }\nonumber\\
&  =\frac{4Te^{2}}{\hbar}\int\frac{d^{3}\mathbf{q}}{\left(  2\pi\right
)  ^{3}%
}\left[  \frac{\gamma_{1}\widehat{v}_{11}^{a}\widehat{v}_{11}^{b}%
}{2\varepsilon_{1}^{3}}+\frac{\gamma_{2}\widehat{v}_{22}^{a}\widehat{v}%
_{22}^{b}}{2\varepsilon_{2}^{3}}+\frac{2\widehat{v}_{12}^{a}\widehat{v}%
_{21}^{b}}{\varepsilon_{1}\varepsilon_{2}\left(  \gamma_{1}^{-1}%
\varepsilon_{1}+\gamma_{2}^{-1}\varepsilon_{2}\right)  }\right].
\label{sigma-general}%
\end{align}
\end{widetext}
with $\widehat{v}_{nk}^{a}\equiv
\sum_{\alpha}\psi_{n,\alpha}\widehat{v}_{\alpha
}^{a}\psi_{k,\alpha}$ (we restored dimensional units in this
formula). Let us stress that the summation here is performed over
the mode indices rather than the band indices in other parts of the
paper.

The main contribution to the paraconductivity comes from the projection to
the singular mode (the first term in the square brackets in Eq.\
(\ref{sigma-general})). Keeping only these terms and using results
(\ref{Eigenvalue1}) for the eigenvalue
$L_{1}^{-1}(\mathbf{q})=\varepsilon_{1}(\mathbf{q})
-i\Omega\gamma_{1}(q_z)$ and (\ref{eigenvector}) for eigenvector, we
derive
\begin{equation}
\sigma_{a}^{\prime}\approx\frac{\pi
e^{2}}{\hbar}\int\frac{d^{3}q}{(2\pi)^{3}
}\frac{\xi_{1,a}^{4}q_{a}^{2}\left[  1+S_{12}r_{a}\tilde{\beta}^{\prime}_{a}\right] ^{2}%
}{\left[  \epsilon+\xi_{1,i}^{2}q_{i}^{2}+S_{12}\tilde{\beta}_{2}\right]
^{3}\,} \label{lastcond}%
\end{equation}
with  $\tilde{\beta}^{\prime}_{1}\equiv 1/\left(
1+\tilde{\theta}\beta_{2}\right)  ^{2}$,
$\tilde{\beta}^{\prime}_{2}\equiv\beta^{\prime}_2/\left(
1+\tilde{\theta}\beta_{2}\right)  ^{2}$ and
$r_{a}=\xi^2_{2,a}/\xi^2_{1,a}$ ($r_z\equiv r$). Below we evaluate the
in-plane and z-axis components of paraconductivity separately.

\subsubsection{In-plane component}

\begin{figure}[ptb]
\begin{center}
\includegraphics[width=3.4in]{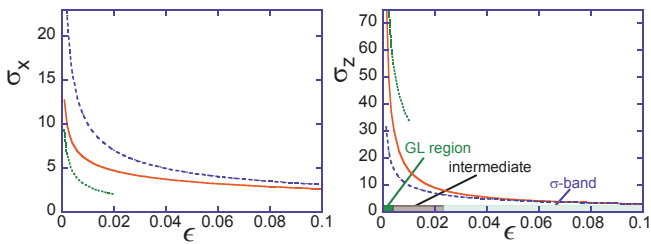}
\end{center}
\caption{Temperature dependencies of the paraconductivity components
(solid lines). \emph{Left panel} shows the in-plane component in units of
$e^{2}/(32\hbar \xi_{z,1})$ (function $\kappa(\epsilon)/\sqrt{\epsilon}$,
see Eqs.\ (\ref{sigma-x}) and (\ref{kappa})). This curve also gives the
fluctuation specific heat in units $1/(8\pi\xi_{1,x}^{2}\xi_{1,z})$, see
Eq.\ (\ref{SpecHeat1}). \emph{Right panel} shows the z-axis component in
units of $e^{2}\xi_{z,1}/(32\hbar\xi_{x,1}^{2})$ calculated from Eq.\
(\ref{sigma-z}). Both components were calculated using parameters
$S_{12}=0.035$, $\widetilde{\theta}=0.377$ and $r=300$. For comparison we
also show the Ginzburg-Landau asymptotics (dotted lines) and $\sigma$-band
contribution to paraconductivity (dashed lines). One can see that in-plane
component diverges slower and z-axis component diverges faster than the
paraconductivity in a single-band anisotropic superconductor. In the
$\sigma_z$-plot we indicate three characteristic regions with different
functional dependencies of fluctuation parameters.}
\label{Fig-sigmas}\end{figure}

In the case of an in-plane paraconductivity we can neglect the small
renormalization of the in-plane velocity in formula
(\ref{lastcond}), perform the $q_{\Vert}$ integration and get%
\[
\sigma_{x}^{\prime}\approx\frac{e^{2}}{32\hbar\pi}\int\limits^{\infty}_{-\infty}
dq_{z}\frac{1}
{\epsilon+\xi_{1,z}^{2}q_{z}^{2}+S_{12}\tilde{\beta}_{2}(q_{z})\,}.
\]
One can see that the in-plane conductivity has exactly the same
temperature dependence as the fluctuation specific heat and
therefore can be represented in the form analogous Eq.\
(\ref{SpecHeat}):
\begin{equation}
\sigma_{x}^{\prime}=\frac{e^{2}}{32\hbar\xi_{1,z}\sqrt{\epsilon}}\kappa(\epsilon),
\label{sigma-x}
\end{equation}
where the function $\kappa(\epsilon)$ is defined by Eq.\ (\ref{kappa}).
The above results show that $\sigma_{x}^{\prime}$ has a 3D character but
diverges slower than the specific $1/\sqrt{\epsilon}$-law. Numerically
calculated dependence $\sigma_{x}^{\prime}\left( \epsilon\right) $ with
parameters typical for MgB$_{2}$, $S_{12}=0.035$,
$\widetilde{\theta}=0.377$ and $r=300$, is shown in the left panel of
Fig.\ \ref{Fig-sigmas} together with the GL asymptotic for $T\rightarrow
T_c$  and the single $\sigma$-band curve. Taking the estimate for the
coherence length of MgB$_{2}$ crystals as \cite{hc2anisotropy}
$\xi_{1,z}\approx2$nm, we find the typical scale for the fluctuation
correction, $\sigma_{x0}^{\prime}=e^{2}/(32\hbar\xi_{1,z})$, as
$\sigma_{x0}^{\prime}\approx$ 40 [$\Omega$ cm]$^{-1}$. This scale must be
much higher in dirty MgB$_{2}$ films. The in-plane paraconductivity in
MgB$_{2}$ films has been studied recently in Ref.\
\onlinecite{SidorJETPLett02} but in the analysis of the experimental data
the two-band nature of MgB$_{2}$ was not taken into account. Nevertheless,
it was found that the paraconductivity indeed diverges slower than
$1/\sqrt{\epsilon}$, in agreement with our results.

\subsubsection{z-component}

The longitudinal component of paraconductivity $\sigma_{z}^{\prime}$
is given by the Eq.\ (\ref{lastcond}). Here the velocity
renormalization turns out to be essential and it cannot be omitted.
Performing integration with respect to $q_{\parallel}$ and using the
introduced above reduced variable $u$, the result can be rewritten
as
\begin{equation}
\sigma_{z}^{\prime}\approx\frac{e^{2}\xi_{1,z}}{32\hbar\xi_{1,x}^{2}\sqrt{\epsilon}}\int_{0}^{\infty}\frac{4du}{\pi}\frac{u^{2}\left[
1+S_{12}r\tilde{\beta}^{\prime}_{2}(u)\right]  ^{2}}{\left[
1+u^{2}+(S_{12}/\epsilon )\tilde{\beta}_{2}(u)\right]  ^{2}\,}.
\label{sigma-z}\end{equation} In the Ginzburg-Landau regime,
$\epsilon\ll1/r+S_{12}$, we obtain\[
\sigma_{z}^{\prime}=\frac{e^{2}\xi_{1,z}\sqrt{1+S_{12}r}}{32\xi_{1,x}^{2}\sqrt{\epsilon}}.
\]

In the regime of nonlocal fluctuations ($\epsilon\gg1/r+S_{12}$),
out of the GL\ region, the behavior $\sigma_{z}^{\prime}$ becomes
quite peculiar. Essential contributions to the integral in Eq.\
(\ref{lastcond}) occur from two regions of $q_{z}$: \ from
$q_{z}\sim1/\xi_{2,z}$ in arguments of functions $\tilde{\beta}_{2}$
and $\tilde{\beta}^{\prime}_{2}$ (mainly $\pi$-band contribution)
and from $q_{z}\sim\sqrt{\epsilon}/\xi_{1,z}$ (mainly $\sigma$-band
contribution). This corresponds to the ranges $u\sim1/r\epsilon\ll1$
and $u\sim1$ in the reduced integral Eq.\ (\ref{sigma-z}). We
evaluate separately contributions from these ranges, or, what is the
same, from $\pi$- and $\sigma$- bands in the most interesting case
$rS_{12}\gg1$ relevant for MgB$_{2}$.

In the range $u\sim1/r\epsilon$ one can drop $\sigma$-band terms
in Eq.\ (\ref{sigma-z}). As we consider the regime $\epsilon\gg
S_{12}$ we can neglect the term with $\tilde{\beta}$ in
denominator. Dimensionalizing the integral, one can obtain the
following result for this term
\begin{align}
\sigma_{z,\pi}^{\prime}  &  \approx\frac{e^{2}\xi_{1,z}S_{12}^{2}\sqrt{r}}{32\hbar\xi_{1,x}^{2}\epsilon^{2}} C_{2}(\tilde{\theta}),\label{sigma-zpi}\\
C_{2}(\tilde{\theta})  &
=\int_{0}^{\infty}\frac{4dv}{\pi}v^{2}\left[
\tilde{\beta}^{\prime}_2\left( v^{2}\right)  \right]
^{2}.\nonumber
\end{align}
Numerical calculation gives $C_{2}(0)\approx1.185$ and
$C_{2}(0.377)\approx0.235$.

In the range $u\sim1$ the main contribution occurs from the $\sigma$-band
and the terms proportional to $S_{12}$ ($\pi$-band terms) can be treated
as small perturbations. Expansion with respect to these terms leads to the
following result for the paraconductivity contribution\begin{equation}
\sigma_{z,\sigma}^{\prime}\approx\frac{e^{2}\xi_{1,z}}{32\hbar\xi_{1,x}^{2}\sqrt{\epsilon}}\left(
1+\frac{S_{12}}{2\epsilon}I_{z}\right) \label{sigma-zsigma}\end{equation}
with\begin{align*} &
I_{z}=\frac{16}{\pi}\int_{0}^{\infty}du\frac{u^{2}}{\left( 1+u^{2}\right)
^{2}}\left[  \epsilon r\ \tilde{\beta}^{\prime}_{2}-\frac{\tilde{\beta}_{2}}{1+u^{2}}\right] \\
&  \approx-\frac{\ln\left(  C_3r\epsilon\right)
}{1+\tilde{\theta}\ln\left( C_4r\epsilon u^{2}\right)  }.
\end{align*}
with $C_3=16\pi^{-2}\gamma_E\exp(-4)\approx 0.053$ and $C_4\approx 2.9$.
Let us note that in this range of temperatures the contribution from
the $\pi$-band comes with the negative sign, as in the cases of the
heat capacity and in-plane paraconductivity. This means that the
$\pi$-band contribution changes sign with increasing temperature.
Comparing contributions (\ref{sigma-zpi}) and (\ref{sigma-zsigma}),
we find that the $\pi$-band term dominates in the interval of
temperatures $\epsilon<S_{12}\left( rS_{12}\right) ^{1/3}$.

Therefore in the case $rS_{12}\gg1\ $ the $z$ axis
paraconductivity $\sigma_{z}^{\prime}$ has three asymptotic
regimes:
\begin{widetext}
\begin{equation}
\sigma_{z}^{\prime}=\frac{e^{2}\xi_{1,z}}{32\hbar\xi_{1,x}^{2}\sqrt{\epsilon
}}\left\{
\begin{array}
[c]{ll}\left(1+S_{12}r\right)^{1/2},&\text{for } \epsilon \ll 1/r+S_{12}\\
\sqrt{r}S_{12}^{2}C_{2}(\tilde{\theta})/\epsilon^{3/2},&\text{for
} 1/r+S_{12}\ll
\epsilon\ll \frac{\left(  rS_{12}\right)  ^{4/3}}{r}\\
1-\frac{S_{12}}{2\epsilon}\frac{\ln\left(  C_{1}r\epsilon\right)
}{1+\tilde{\theta}\ln\left(  C_{1}r\epsilon\right) },&\text{for
}\max[1/r+S_{12}, S_{12}\left( rS_{12}\right) ^{1/3}]\ll\epsilon\ll1
\end{array}
\right.  .
\end{equation}
\end{widetext}
In the case $rS_{12}<1\ $ the intermediate asymptotic disappears.

The right panel of Fig.\ \ref{Fig-sigmas} shows the numerically
calculated dependence of $\sigma _{z}^{\prime}\left( \epsilon\right)
$ with parameters specified in the captions. For comparison, we also
show the GL asymptotics and the $\sigma$-band contribution.

\section{Final remarks}

In this paper we have presented a microscopic derivation
generalizing the conventional GL description onto two-band
superconductors. We have further applied the developed approach to
the investigation of the fluctuation phenomena in MgB$_{2}$. The
important feature of our approach is that the derived
\textit{nonlocal} GL functional takes into account not only the
long-wavelength fluctuations (as is the case of the conventional
GL theory), but also the short-wavelength fluctuations, which
enables to significantly extend the range of validity of the GL
technique. This approach not only permits the study of the
thermodynamic characteristics of the superconductors, but provides
a convenient technique for calculation of transport coefficients.

It is necessary to underline that we succeeded to solve
analytically the problem of accounting for short-wavelength
fluctuations due to the specific for our model large difference in
the intra-band diffusion coefficients $D_\alpha$. Fortunately,
this assumption corresponds to the practically interesting case of
magnesium diboride.

One can see that the main qualitative effect that determines the
unusual behavior of fluctuations in the anisotropic two-band model
consists in globalization of superconductivity in the immediate
vicinity of transition temperature ($\epsilon \ll 1/r+S_{12}$). It
occurs due to the possibility for electrons of both bands to
participate in fluctuation pairing and to exchange of fluctuation
pairs between them.  Formally, this process manifests itself in
the appearance of long range superconducting correlations on the
scale $\widetilde{\xi}_{z}\gg \xi_{1,z}$.

Turning the role of fluctuations in MgB$_{2}$, we refer to the
formulas for paraconductivity in a one-band 3D anisotropic
superconductor \cite{LV02}:
\begin{equation}
\sigma_{x}^{\prime}=\frac{e^{2}}{32\hbar\sqrt{\epsilon}}\frac{\xi_{x}}{\xi_{y}\xi_{z}},\,\,
\sigma_{z}^{\prime}=\frac{e^{2}}{32\hbar\sqrt{\epsilon}}\frac{\xi_{z}}{\xi_{x}\xi_{y}}.
\end{equation}
We see that the growth of the effective $\widetilde{\xi}_{z}$
results in the noticeable increase of the $z$-axis paraconductivity
in the immediate vicinity of critical temperature ($\epsilon \ll
1/r+S_{12}$) with respect to its high temperature ($\epsilon \geq
S_{12}^{4/3}r^{1/3}$) extrapolation based only on the $\sigma$-band
fluctuation pairing. The crossover between these two 3D regimes
occurs in the narrow interval of temperatures $1/r+S_{12}\gg\epsilon
\geq S_{12}^{4/3}r^{1/3}$ where the boson degrees of freedom
corresponding to the pairings in $\pi$-band rapidly freeze out,
leading to the fast decrease ($ \sim \epsilon^{-2}$) of $z$-axis
paraconductivity. We want to stress that this temperature dependence
appears due to the noticeable contribution of short-wavelength
fluctuations in this range of temperatures.

In the case of the in-plane component of paraconductivity and the
fluctuation part of heat capacity, the situation is the opposite. The
coherence length $\xi_{z}$ appears in the denominator of the corresponding
expressions, so $\sigma_{x}^{\prime}$ and $C'$ turn out to be suppressed
in the vicinity of transition with respect to their extrapolation formulas
from the high-temperature behavior. Similar situation occurs in the
temperature dependencies of the in-plane upper critical field and of
anisotropy of the upper critical field \cite{GK03a,GK03b}.

The obtained results coincide with that ones of the diagrammatic approach
but the proposed GL description has an advantage of being more physically
transparent, economic, and universal. The fact that the results derived
via the GL equations coincide with those of our microscopic consideration
is a convincing cross-check of the phenomenological description. The same
approach can be used to derive other fluctuation properties such as the
magnetic susceptibility and the field dependencies of conductivity and
magnetization in the vicinity of transition.

\section{Acknowledgements}

This work was supported by the U.S. DOE, Office of Science, under
contract \# W-31-109-ENG-38. A.A.V. acknowledges the support of the
FIRB project of the Italian Ministry of Science and Education.


\begin{thebibliography}{99}
\bibitem{GL50}V.\ L.\ Ginzburg and L. D. Landau, Zh.\ Eksp.\ Teor. Fiz.\ ,
\textbf{20}, 1064(1950).

\bibitem{AbrikosovJETP57}A. A. Abrikosov, Zh.\ Eksper.\ i
Teor.\ Fiz.\ \textbf{32}, 1442(1957) (Sov.\ Phys.\ JETP \textbf{5},
1174(1957)).

\bibitem{BGFLV94}G.\ Blatter, M.\ V.\ Feigel'man, V.\ B.\ Geshkenbein,
A.\ I.\ Larkin, and V.M.Vinokur, Rev. of Mod. Phys., \textbf{66},
1180(1994).

\bibitem{F03}L.Viverit, G.M.Brun, A.Minguzzi, and R. Fazio, \textit{Pairing
fluctuations in trapped Fermi gases}, cond-mat/0402620 v1 (2004).

\bibitem{G58}L.\ P.\ Gor'kov, Zh.\ Eksp.\ Teor. Fiz.\ \textbf{36}, 1918,
(1959) [Sov.\ Phys.\ JETP, \textbf{9}, 1364 (1959)]; L.\ P.\ Gor'kov and T.\
K. Melik-Barkhudarov, Zh.\ Eksp.\ Teor. Fiz.\ \textbf{45}, 1493, (1963)
[Sov.\ Phys.\ JETP, \textbf{18}, 1031 (1964)].

\bibitem{GK03a}A.\ A.\ Golubov and A.\ E.\ Koshelev, Phys.\ Rev.\ B
\textbf{68}, 104503 (2003).

\bibitem{GK03b}A.\ E.\ Koshelev and A.\ A.\ Golubov, Phys.\ Rev.\ Lett.
\textbf{92}, 107008 (2004).

\bibitem{hc2anisotropy}Yu.\ Eltsev, S. Lee, K. Nakao, N. Chikumoto, S.
Tajima, N. Koshizuka, and M. Murakami, Phys.\ Rev.\ B \textbf{65}, 140501(R)
(2002); Physica C \textbf{378-381}, 61 (2002). L.\ Lyard, P. Samuely, P.
Szabo, T. Klein, C. Marcenat, L.Paulius, K.H.P. Kim, C.U. Jung, H.-S. Lee,
B. Kang, S. Choi, S.-I. Lee, J. Marcus, S. Blanchard, A.G.M. Jansen, U.
Welp, G. Karapetrov, W.K. Kwok., Phys.\ Rev.\ B \textbf{66}, 180502(R)
(2002); M.\ Angst, R.\ Puzniak, A.\ Wisniewski, J.\ Jun, S.\ M.\ Kazakov,
J.\ Karpinski, J.\ Roos, and H.\ Keller, Phys.\ Rev.\ Lett.\ \textbf{88},
167004 (2002).

\bibitem{AGD}A.\ A.\ Abrikosov, L.\ P.\ Gorkov, and I.\ E.\ Dzyaloshinski,
\textit{Methods of quantum field theory in statistical physics},
Prentice-Hall, Inc., Englewood Cliffs, NJ, 1963.

\bibitem{Multi-Two}Strictly speaking, the band structure of MgB$_{2}$ is
composed of four bands: two $\sigma-$bands and two $\pi$-bands (see, e.g.,
Ref.\ \onlinecite{LiuPRL01}). However, in real samples the scattering rate
between the bands of the same family is usually high while the scattering of
electrons between the different families is rather weak.\cite{M03} In this
situation scattering homogenizes superconductivity inside both band families
and superconductivity in magnesium diboride is effectively described by the
two-band model. To simplify terminology, in the rest of the paper we use
term ``$\sigma$-band'' (``$\pi$-band'') for the family of $\sigma$-bands
($\pi$-bands).

\bibitem{LV02}A.\ Larkin and A.\ Varlamov, \textit{Theory of fluctuations in
superconductors}, Oxford University press (2004).

\bibitem{E61}G.\ M.\ Eliashberg,   Zh.\ Eksp.\ i Teor.\ Fis.\ \textbf{41},
1241 (1961) [ Sov.\ Phys.\ - JETP , \textbf{14}, 856 (1961)].

\bibitem{highHc2}
R.\ H.\ T. Wilke, S.\ L.\ Bud'ko, P.\ C.\ Canfield, D.\ K.\ Finnemore, R.\
J.\ Suplinskas, and S.\ T.\ Hannahs,Phys.\ Rev.\ Lett.\ \textbf{92},
217003 (2004); V.\ Braccini, \textit{et al.}
cond-mat/0402001; M.\ Angst, S.\ L.\ Bud'ko, R.\ H.\ T.\ Wilke, P.\ C.\
Canfield cond-mat/0410722



\bibitem{M03}I.\ I.\ Mazin, O.\ K.\ Andersen, O.\ Jepsen, O.\ V.\ Dolgov,
J.\ Kortus, A.\ A.\ Golubov, A.\ B.\ Kuz'menko, and D.\ van der Marel,
Phys.\ Rev.\ Lett.\ \textbf{89} 107002 (2002).

\bibitem{Suhl}H.\ Suhl, B.\ T.\ Matthias, and L.\ R.\ Walker,
Phys.\ Rev.\ Lett.\ \textbf{3}, 552 (1959).

\bibitem{Moskal}V.\ A.\ Moskalenko, Fiz. Met. Met. \textbf{4}, 503 (1959).

\bibitem{Kortus}J.\ Kortus, I.\ I.\ Mazin, K.\ D.\ Belashchenko,
V.\ P.\ Antropov, L.\ L.\ Boyer, Phys. Rev. Lett. \textbf{86}, 4656 (2001).

\bibitem{An}J.\ N.\ An and W.\ E.\ Pickett, Phys.\ Rev.\ Lett.\ \textbf{86},
4366 (2001).

\bibitem{LiuPRL01}A.\ Y.\ Liu, I.\ I.\ Mazin, and J.\ Kortus,
Phys.\ Rev.\ Lett.\ \textbf{87}, 087005 (2001).

\bibitem{Kong}Y.\ Kong, O.\ V.\ Dolgov, O.\ Jepsen, and O.\ K.\ Andersen ,
Phys. Rev. B \textbf{64}, 020501(R) (2001).

\bibitem{Yildirim}T.\ Yildirim, O.\ G\"{u}lseren, J.\ W.\ Lynn,
C.\ M.\ Brown, T.\ J.\ Udovic, Q.\ Huang, N.\ Rogado, K.\ A.\ Regan, M.\ A.\
Hayward, J.\ S.\ Slusky, T.\ He, M.\ K.\ Haas, P.\ Khalifah, K.\ Inumaru,
and R.\ J.\ Cava , Phys. Rev. Lett. \textbf{87}, 037001 (2001).

\bibitem{Choi}H.\ J.\ Choi, D.\ Roundy, H.\ Sun, M.\ L.\ Cohen, and
G.\ Louie, Phys.\ Rev.\ B, \textbf{66}, 020513 (2002); Nature \textbf{418},
758 (2002).

\bibitem{Golub}A.\ A.\ Golubov, J.\ Kortus, O.\ V.\ Dolgov, O.\ Jepsen,
Y.\ Kong, O.\ K.\ Andersen, B.\ J.\ Gibson, K.\ Ahn, and R.\ K.\ Kremer,
J.Phys.: Condens. Matter, \textbf{14}, 1353 (2002).

\bibitem{MakiHc2}K.\ Maki, Physics, \textbf{1}, 21 (1964).

\bibitem{CarringtonPhysC03}A.\ Carrington and F.\ Manzano,
Physica C, \textbf{385}, 205 (2003).

\bibitem{Kurtus04} J.\ Kortus, O.\ V.\ Dolgov, R.\ K.\ Kremer, and A.\ A.\
Golubov,
cond-mat/0411667

\bibitem{SidorJETPLett02}A. S. Sidorenko, L. R. Tagirov, A. N. Rossolenko, N.
S. Sidorov, V. I. Zdravkov, V. V. Ryazanov, M. Klemm, S. Horn, and R.
Tidecks, JETP Letters \textbf{76}, 17 (2002).
\end{thebibliography}
\end{document}